# Development of large radii half-wave plates for CMB satellite missions

G. Pisano*[a,b], B. Maffei[b], M.W. Ng[b], V. Haynes[b], M. Brown[b], F. Noviello[b], P. de Bernardis[c], S. Masi[c], F. Piacentini[c], L. Pagano[c], M. Salatino[c], B. Ellison[d], M. Henry[d], P. de Maagt[e], B. Shortt[f]

[a]School of Physics and Astronomy, Cardiff University, Queen's Building, The Parade, CF24 3AA Cardiff (United Kingdom); [b]School of Physics and Astronomy, The University of Manchester, Oxford Road, M13 9PL Manchester (United Kingdom); [c]Dipartimento di Fisica, Universita' di Roma La Sapienza, P.le A.Moro 5, 00185 Roma (Italy); [d]Rutherford Appleton Laboratory, Building R68, Harwell Oxford, Didcot, Oxfordshire, OX11 0QX (United Kidgdom); [e]Electromagnetics & Space Environments Division, European Space Agency, NL 2200 AG Noordwijk, The Netherlands; [f]Future Missions Office, Science Directorate, European Space Agency, NL 2200 AG Noordwijk, The Netherlands;

**ABSTRACT**

The successful European Space Agency (ESA) Planck mission has mapped the Cosmic Microwave Background (CMB) temperature anisotropy with unprecedented accuracy. However, Planck was not designed to detect the polarised components of the CMB with comparable precision. The BICEP2 collaboration has recently reported the first detection of the B-mode polarisation. ESA is funding the development of critical enabling technologies associated with B-mode polarisation detection, one of these being large diameter half-wave plates. We compare different polarisation modulators and discuss their respective trade-offs in terms of manufacturing, RF performance and thermo-mechanical properties. We then select the most appropriate solution for future satellite missions, optimized for the detection of B-modes.
**Keywords:** Cosmic Microwave Background, Satellite, Optics, Polarimetry, Half-Wave plate.

## 1. INTRODUCTION

The Cosmic Microwave Background (CMB) radiation encodes a wealth of information about the Universe and its evolution. This radiation was last scattered with matter 380'000 years after the Big Bang, and travelled freely afterwards. Since it was emitted when the Universe was very uniform and in thermal equilibrium, it has a Black-Body spectrum and it is extremely isotropic, with a level of anisotropy of the order of $10^{-5}$. Statistical distribution of this anisotropy is directly related to the values of the so called cosmological parameters, a set of parameters that describe the evolution of the Universe, its metric, and its constituents. The CMB radiation also presents a small level of polarisation, which can be decomposed in two fields with different symmetry properties: the E-mode, and the B-mode. The latter mode is of particular interest for cosmology, since it is a univocal signature of gravitational waves in the early Universe. The first detection of this signature has been reported by the BICEP2 South Pole based instrument [1]. The B-mode signal claimed in the BICEP2 results is at the level of $10^{-7}$ or below the CMB intensity. Detecting such signal requires extreme polarisation purity and control of systematic effects, which are at the basis of the design of any next generation CMB experiment, ground-based, balloon-based, or satellite based.

The goal of this project is to develop large diameter HWPs, of the order of ~1.2m $\varnothing$, with the aim of defining a road map for the subsequent development of larger waveplates (up to ~1.7m $\varnothing$) as required by a COrE-like satellite [2]. This implies a radical increase in dimensions from the presently available waveplates (~30cm) by a factor of four in diameter and sixteen in area. We summarise the concept design selection carried out in the ESA-ITT 'Large radii HWP development' study. In Section 2 we review the different polarisation modulation techniques used at millimetre wavelengths, limiting our discussion only to the quasi-optical modulators, relevant for this work. Section 3 is dedicated to the trade-off between the different techniques leading to the final concept design selection of the most appropriate polarisation modulator.

*giampaolo.pisano@astro.cf.ac.uk; phone ++44(0)29208 76460.

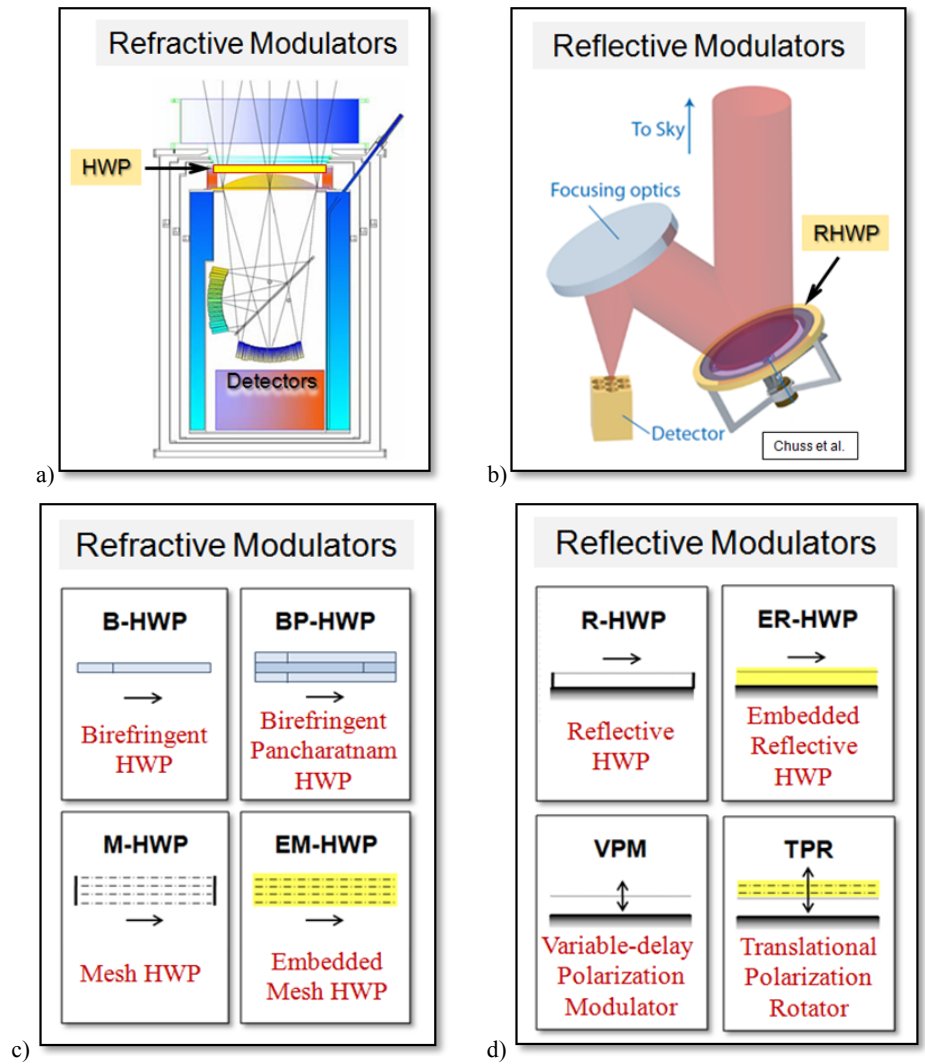

Figure 1. Polarisation modulators types (top left and top right figures respectively from [4] and [5]).

## 2. REVIEW OF QUASI-OPTICAL POLARIZATION MODULATION TECHNIQUES

The quasi-optical modulation of linear polarisation can be achieved in a number of different ways involving electro-mechanical techniques. These involve devices ranging from simple rotating polarisers, K-mirrors and Fresnel-double rhombs [3], to more sophisticated components that will be discussed in this paper. We can divide polarisation modulators in two categories, which are *refractive* and *reflective* modulators. These classifications depend upon whether the modulated radiation passes through the device or is reflected from it (see Fig.1). Ideally the modulator should be the first optical element of an instrument, in order to avoid modulation of the spurious polarisation induced by other elements of the optical chain. It is also preferable for the modulator to be cryogenically cooled in order to reduce its losses and associated thermal emissions, which would most likely compromise the sensitivity of the instrument detectors.

In the refractive system example shown in Fig.1a, a rotating HWP is positioned inside a cryostat before a focusing lens, whereas in the reflective system of Fig.1b, the reflecting HWP is positioned before a focusing mirror. Depending on the specific optical configuration, cryostat windows and thermal filters might be required before the modulator elements.

Within the refractive modulator category we discuss four different Half Wave Plate designs, sketched in Fig.1c:

- Birefringent single plate Half Wave Plate (B-HWP)
- Birefringent multi-plate Pancharatnam HWP (BP-HWP)

- Air-gap Mesh HWP (M-HWP)
- Dielectrically Embedded Mesh HWP (EM-HWP)

With regard to the reflective modulator category, we discuss the designs sketched in Fig.1d:
- Air-gap Reflective HWP (R-HWP)
- Embedded Reflective HWP (ER-RHWP)
- Variable-delay Polarisation Modulator (VPM)
- Translational Polarization Rotator (TPR)

We will consider the relative advantages and disadvantages of each concept as well as open issues associated with them. We anticipate that, due to stringent manufacturing limits associated with some of these devices, we will be able to pre-select three options. Other considerations, from the RF performance and the cryo-mechanical points of view, will allow us to select what we think would be the most appropriate solution for future large diameter B-mode satellite instruments.

## 2.1 Birefringent single plate Half Wave Plate (B-HWP)

The modulation of a linearly polarised signal can be achieved by directing the signal source towards a rotating HWP and then detecting the corresponding transmitted component with a polarisation sensitive detector. The simplest HWP design consists of a birefringent plate cut with the crystal birefringent axis parallel to its faces. The difference in refractive index between the ordinary and extraordinary axes lying on the plate will create a phase-shift between two orthogonal polarisations parallel to them. The plate thickness is designed to provide the required 'half-wave' differential phase-shift of 180° (Fig.2).

In a rotating HWP, the combined effect of the mechanical rotation and the 180° differential phase-shift rotates the polarisation vector of an incoming linearly polarised signal at twice the mechanical angular velocity. Using a linear polariser the intensity of the signal will be modulated sinusoidally in intensity before reaching the detector. Interestingly, in the case of an ideal HWP, any unpolarised signal will not be modulated as the phase of the orthogonal polarisations is not correlated (Fig.2). There are systematic effects introduced by this type of device that have been theoretically investigated in different optical configurations [6].

Critical issues:

- A single plate HWP is inherently a narrow-band device because the exact differential 180° phase-shift occurs only at specific frequencies related to the plate thickness.
- At millimetre wavelengths there are not many birefringent materials available. The most commonly used material is sapphire, which has a high refractive index (n~3.3). This creates a strong mismatch with plane waves propagating in free space, resulting in high reflection coefficients and the need for anti-reflection coatings (ARCs).
- ARCs are materials with refractive index of the order of $\sqrt{n}$. In the case of sapphire, these materials are not available naturally and they have to be artificially synthesised.

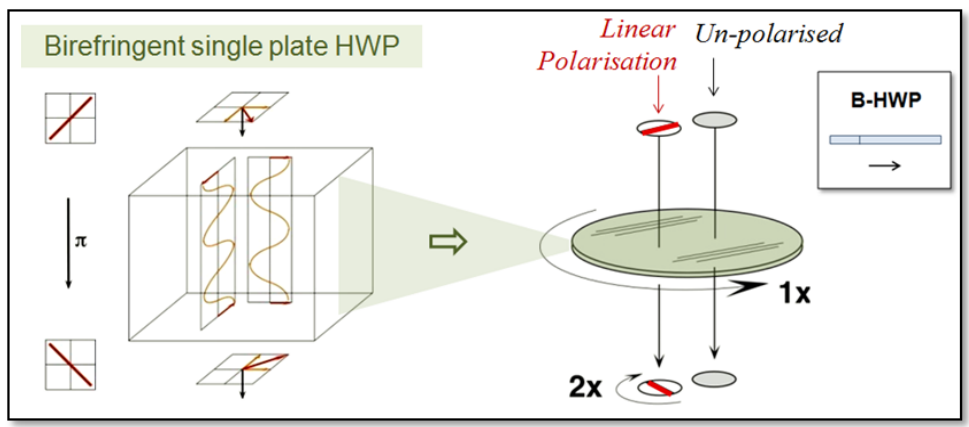

Figure 2. Working principle of a birefringent HWP. (Figures from [3]).

## 2.2 Birefringent multi-plate Pancharatnam HWP (BP-HWP)

It is possible to increase the bandwidth of a waveplate using the theory developed by Pancharatnam in the 50s. The multi-plate method suggested involves the stacking of different birefringent HWPs and rotating their axes by specific angles. It is then possible to achieve very broadband performance (Fig.3). There are different 'recipes' with increasing number of plates and relative bandwidth. A prototype of a 3-plate Pancharatnam recipe is shown in Fig.3. Detailed studies have been carried out in order to model the performance and the systematic effects of these devices. As an example, Fig.4 shows the accurate agreement between modelling and experimental data of a 5-plate sapphire HWP [7,8].

Critical issues:

- To avoid surface mismatch, and consequent performance limitations, very broadband multi-layer ARCs are required.
- There is a well know problem of delamination of the AR coatings when the devices are subject to repeated cryogenic cooling cycles. Artificial dielectrics based on mesh technology have been adopted as a potential solution [9].
- Common also to the previous device is the limited availability of large diameter crystal plates. The maximum commercially available diameter is of the order of ~30 cm. This implies that, at the present moment, the birefringent HWPs are not suitable for large diameter arrays applications.

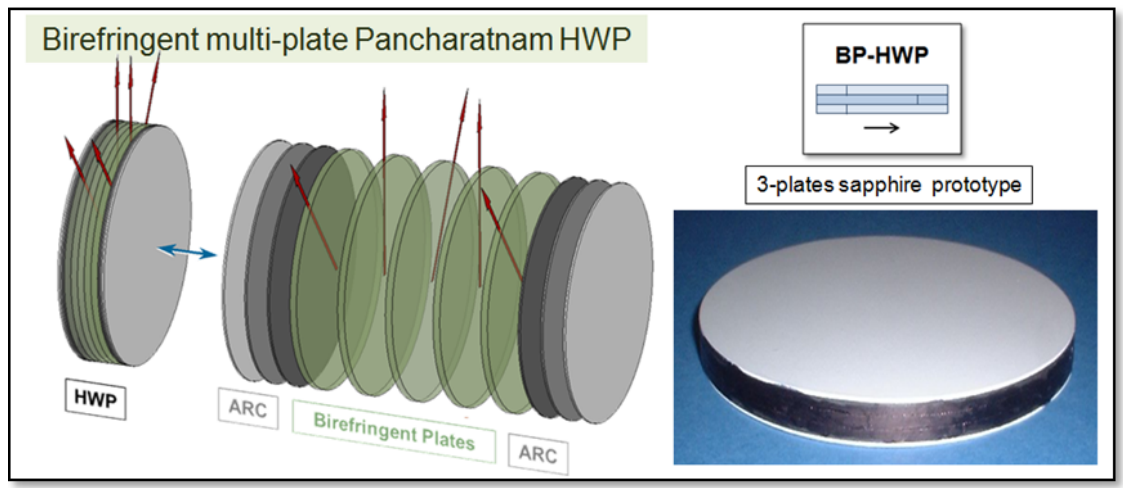

Figure 3. Broadband HWPs made with combinations of birefringent plates (left figure from [8]).

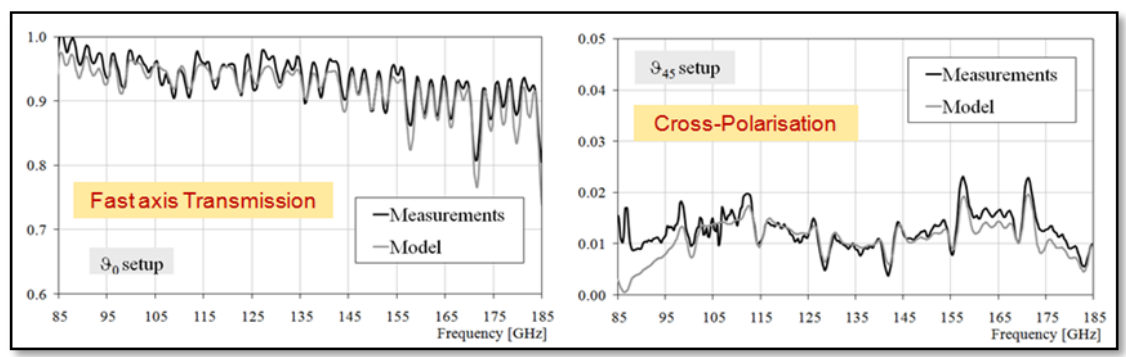

Figure 4. Modelling and measurements of a 5-plate Pancharatnam HWP (plots from [7]).

## 2.3 Air-gap Mesh HWP (M-HWP)

The photo-lithographic techniques used to make mesh filters at millimetre wavelengths have also been adopted to manufacture HWPs [5, 10]. This concept uses 'anisotropic' mesh grids that possess a variation in performance related to the two orthogonal polarisation directions. It is possible to design grids that exhibit capacitive or inductive behaviour in one polarisation direction and that are almost transparent in the orthogonal one. The *Pol-1* signal in Fig.5 interacts with the first three capacitive grids and then propagates through the remaining inductive grids almost unaffected. Similarly, *Pol-2* does not respond to the capacitive grids, but instead interacts with the inductive ones. The capacitive and inductive grids introduce phase shifts with opposite signs making it possible to achieve large differential phase shifts between the orthogonal polarisations. A picture of a prototype is presented in Fig.5 whereas Fig.6 shows the agreement between experimental data and modelling [10]. The metal grids are obtained evaporating copper on thin dielectric substrates and then etching away the desired material using photolithography.

Critical issues:

- Although these devices can be used at cryogenic temperatures, this technique cannot be used for large diameter devices due to the fragility of the thin dielectric substrates (few μm) necessary to hold the metal grids.

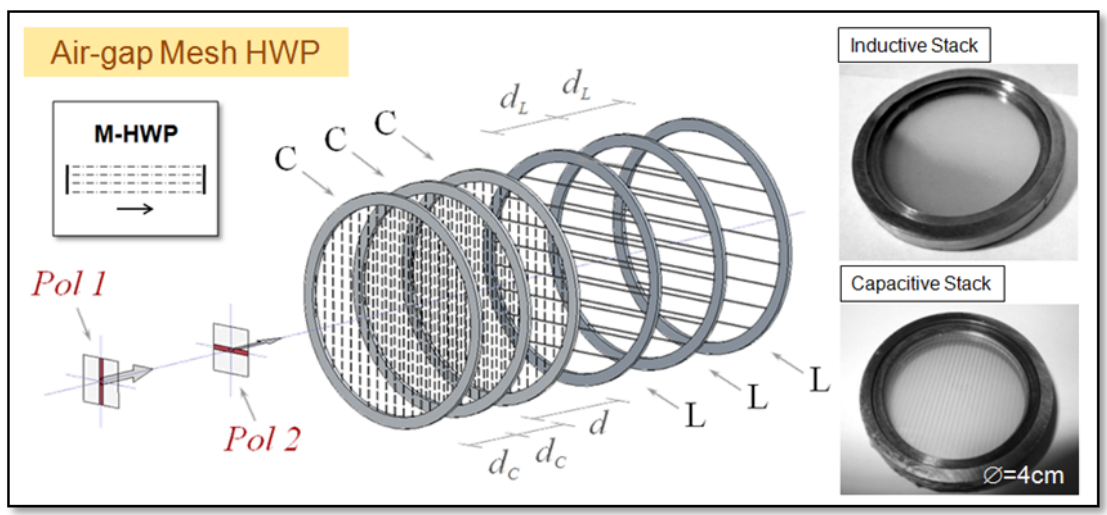

Figure 5. First broadband mesh-HWP made with free standing metal mesh grids (figures from [10]).

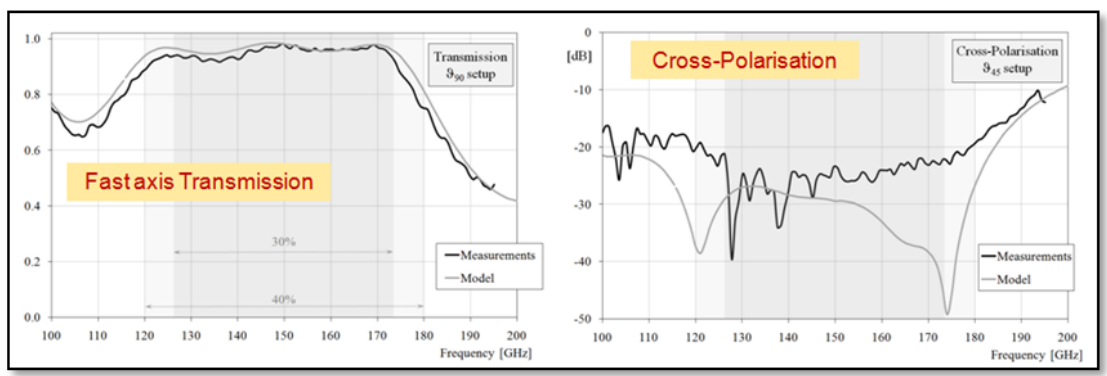

Figure 6. Examples of modelling and measurements of an air-gap mesh-HWP (plots from [10]).

## 2.4 Dielectrically Embedded Mesh HWP (EM-HWP)

The fragility problem of the air-gap mesh HWP can be solved embedding all the grids inside the same dielectric material. The grid-substrates are now thermally bonded with other layers of the same material that function as spacers. This technique has been largely used in the past for the manufacture of cryogenic mesh-filters [11]. The development of these devices led to designs with grid geometries with both the capacitive and inductive lines on the same grid (Fig.7). In this case the device thickness was reduced by a factor two and the losses brought down to ~1% and ~2% for the C- and L- axis respectively, at room temperature. The performance of a W-band mesh-HWP prototype is reported in Fig.8 [12]: for the first time the differential phase-shift was measured directly with a Vector Network Analyser and not inferred from intensity measurements. Experimental investigations of the impact on the beam induced by a dielectrically embedded mesh-HWPs (and mesh-QWP) have been also carried out [13, 14]. A lot of progress was achieved recently in the design of EM-HWPs leading to bandwidths increase from the initial 25-30% up to ~100%.

Critical issues:

- Although the measured losses are close to those expected by the modelling there is an expected slight difference in absorption between the waveplate axes with the inductive one exhibiting higher losses.

- The potential gradient in temperature across large plates of this type needs to be investigated.

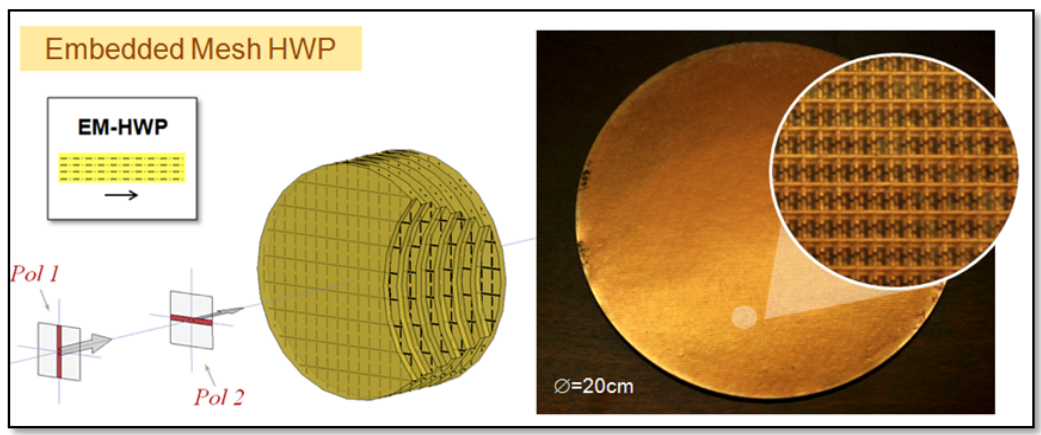

Figure 7. W-band embedded mesh-HWP made at JBCA, the University of Manchester (figures from [12], reproduced courtesy of The Electromagnetics Academy).

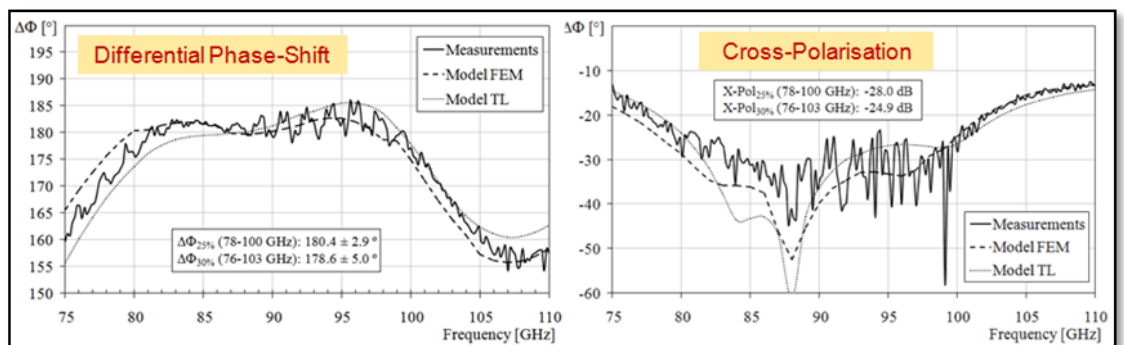

Figure 8. W-band embedded mesh-HWP phase and cross-polar performance (plots from [12], reproduced courtesy of The Electromagnetics Academy).

## 2.5 Air-gap Reflective HWP (R-HWP)

A Reflecting Half-Wave Plate (R-HWP) consists of a free-standing wire grid polariser located at a specific distance from a flat mirror (Fig.9). The R-HWP rotates around the axis orthogonal to its surface and passing through its centre. Assuming the R-HWP is oriented as in Fig.9, depending on the angle of incidence φ (dictated by the optical setup), the R-HWP will introduce a phase-shift between the two orthogonal 's' and 'p' polarisations, respectively orthogonal and parallel to the plane of incidence. The *s-pol* signal will be reflected by the wire-grid whereas the *p-pol* will pass through it and will then be reflected by the mirror (the opposite will happen when the wire-grid is rotated by 90°). When the phase-shift is equal to 180° and the R-HWP rotates with angular velocity ω the overall effect is a rotation of the reflected linear polarisation angle at a frequency 2ω. This happen only when the path difference between the two waves is equal to π, i.e. at the following frequencies:

$$\nu_n = (2n+1)\nu_0 \quad \text{where} \quad \nu_0 = \frac{c}{4d\cos\varphi} \tag{1}$$

where *d* is the distance between the wire-grid and the mirror. This implies that the R-HWP polarisation modulation efficiency vs. frequency has a sinusoidal behaviour with maxima at frequencies $\nu_n$ (see example in Fig.10). If no spectral filtering is adopted, the average modulation efficiency across large bandwidths, is of the order of 0.5. Adding spectral filtering with periodic narrow bands it is possible to achieve higher efficiencies.

Critical issues:

- The modulation efficiency sinusoidal behaviour implies that it can be used only within periodic narrow bands, where the differential phase-shift is close (and equal) to 180°.

- The free-standing wire-grid is very fragile and the manufacture of large diameter devices is challenging. Moreover, the grid-mirror distance has to be constant across the plate and this is very difficult to achieve with large diameter plates within the required accuracy.

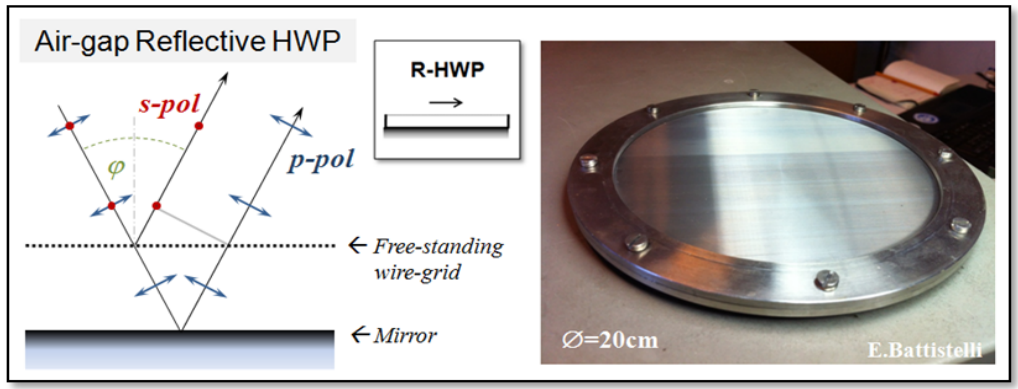

Figure 9. Reflective HWP concept and prototype built at the Rome University (photograph from E.Battistelli).

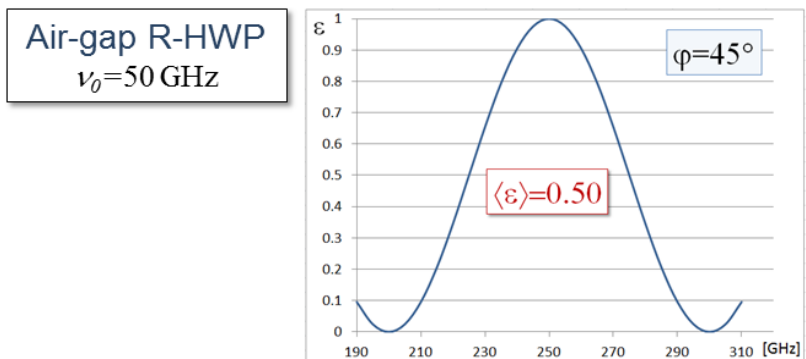

Figure 10. Example of polarization modulation efficiency of an air-gap Reflective HWP.

## 2.6 Embedded Reflective HWP (ER-RHWP)

It is possible to address the critical issues of the air-gap R-HWP by designing dielectrically embedded versions of it. Two types of Embedded Reflective HWPs have been investigated (Fig.9):

- ER-HWP-A: based on photolithographic wire-grid and dielectric substrates;
- ER-HWP-B: based on metal mesh metamaterial substrates.

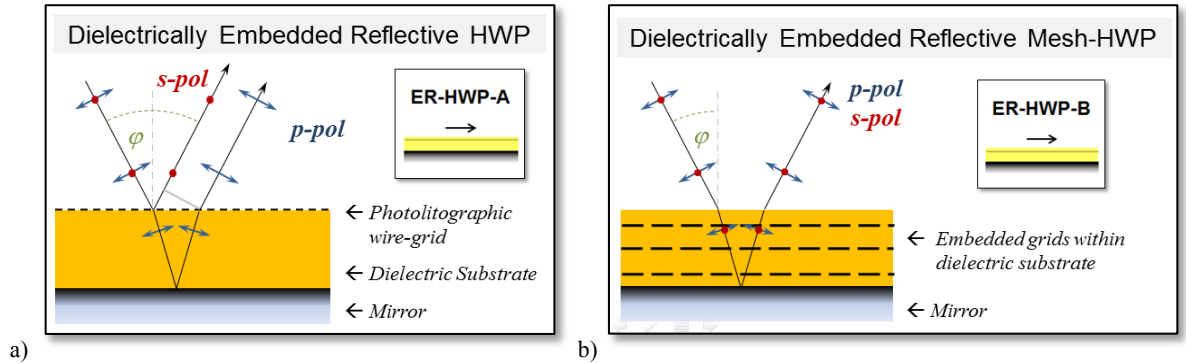

Figure 11. Dielectrically embedded reflective HWP concepts.

In the first case, ER-HWP-A, a dielectric substrate can be used to replace the air-gap and a photolithographic wire-grid can be processed directly on the substrate, avoiding building large diameter free-standing wire-grids (Fig.11a). This is a very robust device with a grid-mirror distance constant down to the required accuracy. However, both the mismatch introduced by the substrate and the off-axis operating conditions deteriorates the polarization modulation efficiency. It is possible to recover this efficiency loss by using more dielectric layers obtaining "top-hat" responses around the periodic peaks. An example is shown in Fig.12 where the off-axis efficiency varies between 53-58% depending on the wire-grid orientation with respect to the plane of incidence ($\alpha=0$, 90° corresponds to wires respectively perpendicular and parallel to that plane). We notice that this device has a higher performance than the one achievable with the classic air-gap version.

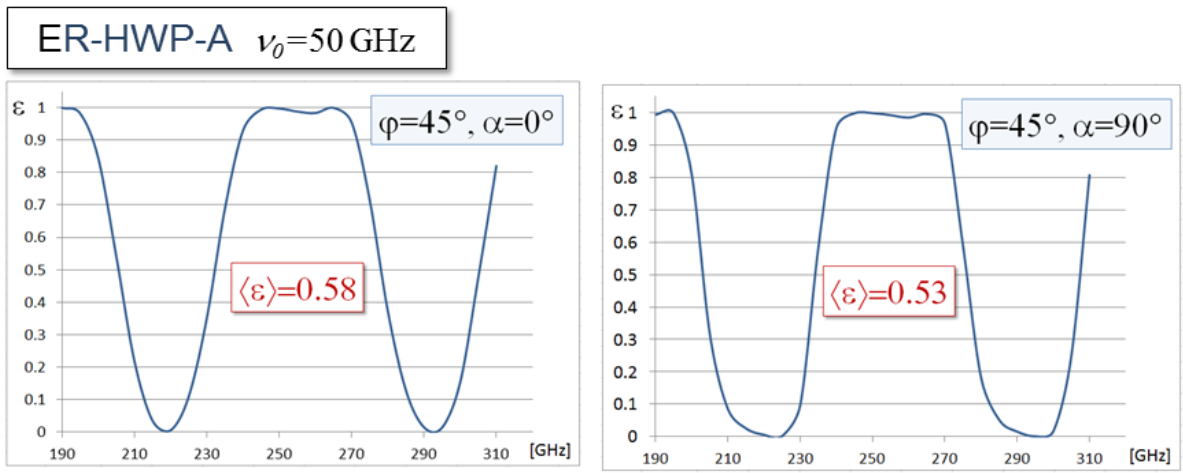

Figure 12. Dielectrically embedded ER-HWP-A polarization modulation off-axis example.

The second solution, the ER-HWP-B, is based on artificial dielectric layers (Fig.11b). These metamaterials with custom-designed permittivities can be realised adopting the same embedded mesh technology used to manufacture the other devices discussed earlier [9]. Fig.13 shows an example of on-axis performance of an ER-HWP-B type device. Average efficiencies around 75% and 'top-hat' responses over very large bandwidths, of the order of ~150%, can be achieved even in off-axis conditions. We notice that this type of HWP provides the largest bandwidth achievable within all the technologies discussed in this paper.

Critical issues:

- The ER-HWP-A device still shows periodic narrow bands and additional filtering would be required.

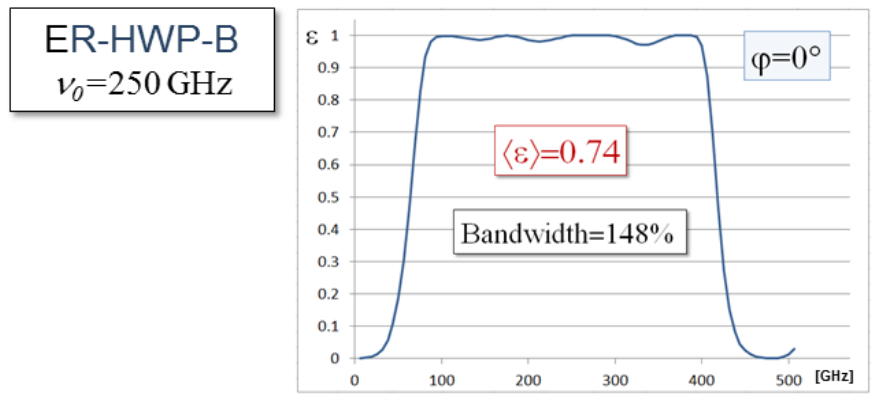

Figure 13. Dielectrically embedded ER-HWP-B polarization modulation on-axis example.

## 2.7 Variable-delay Polarisation Modulator (VPM)

The Variable-delay Polarisation Modulator (VPM) is based on the introduction of a phase delay between two orthogonal linear polarizations [5, 15]. The linear polarisation states are separated by means of a wire-grid polariser placed parallel to a metallic mirror (Fig.14). The polarisation modulation is achieved by changing the separation between the polarising grid and the parallel mirror. The VPM modulates only the phase of the field component orthogonal to the polariser wires.

Critical issues:

- Because the phase-shift is approximately proportional to the additional optical path, and depends also on the frequency of the radiation, the VPM is inherently a narrow-band device.
- The VPM does not modulate the Stokes parameters Q & U at the same time, unlike a rotating HWP. Depending on the orientation of the grid it modulates either Q & V or U & V which results in the linear polarisation being converted back and forth from linear to circular, but with no vector rotation.

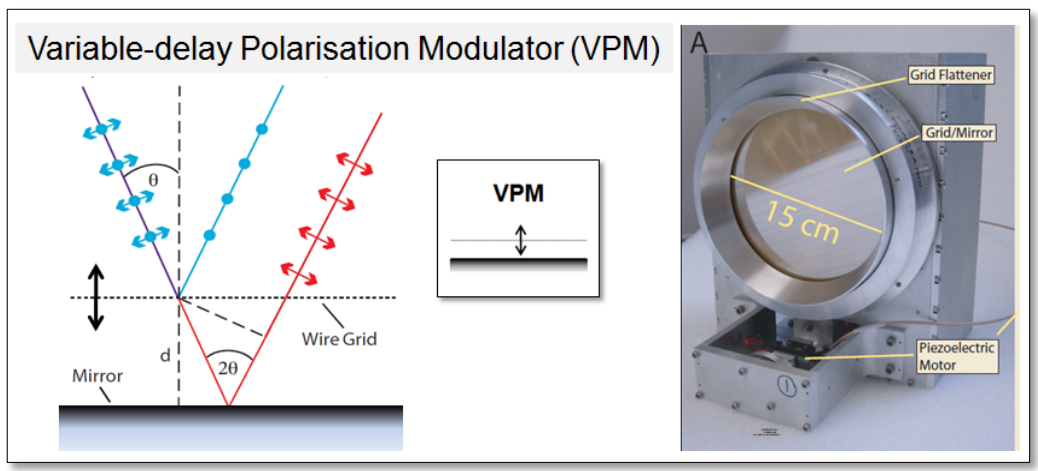

Figure 14. Variable-delay Polarisation Modulator designed and built at NASA GSFC (figures from [15], © IOP Publishing. Reproduced by permission of IOP Publishing. All rights reserved).

## 2.8 Translational Polarization Rotator (TPR)

The modulation problem of the VPM can be solved if the variable phase is introduced between left- and right- handed circular polarisations rather than linear polarisations (Fig.15). In the Translational Polarization Rotator [16] the circular polarisation states of the incoming radiation are separated by a circular polarizer that is made combining a Quarter Wave Plate (QWP) and a wire-grid, oriented at 45° from each other. A movable mirror is positioned behind and parallel to the circular polariser and, as its distance varies, an incident linear polarisation is rotated through an angle that is proportional to the introduced phase delay. The device modulates both the Q & U Stokes parameters similar to a rotating HWP. The QWP used was based on the embedded mesh technology.

Critical issues:

- The TPR performance is limited by the mesh-QWP bandwidth and losses that are similar to those discussed in the EM-HWP case. However, the number of grids of the mesh-QWP is reduced by a factor 2 compared to a mesh-HWP.
- The gradient in temperature across large plates needs to be investigated.

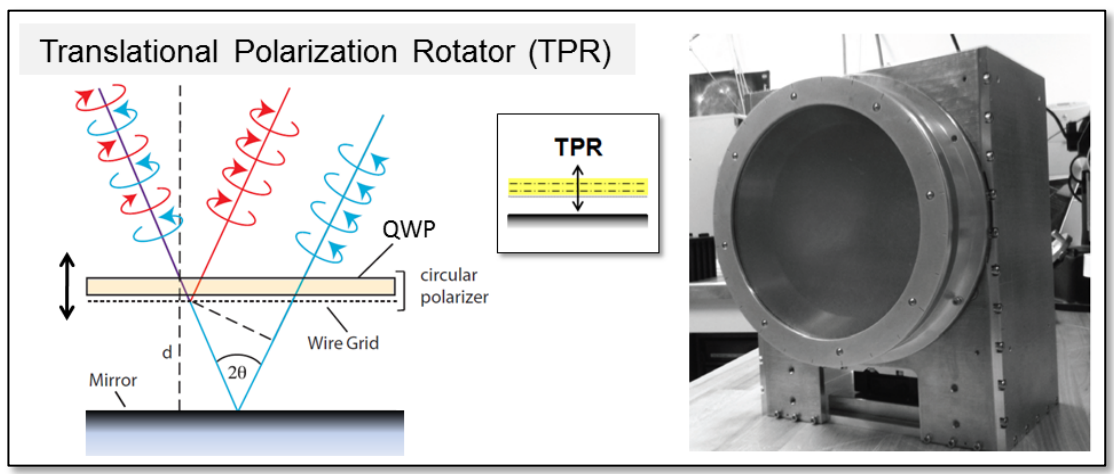

Figure 15. TPR modulator built at NASA-GSFC (figures from [16]).

## 3. TRADE-OFF DISCUSSION AND CONCEPT DESIGN SELECTION

The aim of this study is the development of large radii half-wave plates with respect to their relevance for observational cosmology and the specific application of mapping the cosmic microwave background polarization from space. We have presented eight different concepts for polarisation modulation and a summary review of their respective performance and relative merits. We consider now that there are three fundamental aspects of HWP modulator design that are key to achieving the required experimental performance in spaceflight conditions. These are:

- HWP manufacturability
- HWP Radio Frequency (RF) performance
- HWP suitability for cryogenic and mechanical operation in a space environment

We believe that these key attributes are common and relevant selection criteria for all of the concepts presented and their respective investigation is necessary within this design selection. With these key factors evaluated and assessed, we will then be able to move forward with a final device concept selection. We discuss our considerations regarding each of the key attributes in the following sections.

## 3.1 HWP manufacturability

Critical to manufacture is the identification and availability of suitable materials that are compliant with spaceflight requirements, e.g. low mass, low outgassing, launch survivability; have adequate RF performance and possess sufficient mechanical robustness to maintain appropriate mechanical tolerances during launch and normal spaceflight operation. The trade-off Table 1 provides the characteristics of the eight modulators identifying their relevant advantages and disadvantages. The table is divided into categories of transmission and reflection modulators, mechanical-rotation or translational-motion, and birefringent materials or metal grids. The meaning of the colours used in this and the following tables is the following: a green box means an 'advantage' of that device; a red box means a 'not usable' device; a grey box means a 'disadvantage', but still a usable device; a white box means a parameter not available/applicable; a yellow box means the 'goal' of this study.

Due to inherent material limitations we are of the opinion that HWP modulators constructed from birefringent materials are unsuitable candidates because of the present unavailability of birefringent crystals with diameters larger than ~ 30cm. We expect that both the single plate (B-HWP) and the multi-plate Pancharatnam (BP-HWP) birefringent concepts will not be considered further as options, though we will include a description of their attributes and current performance as a part of the conceptual design selection task.

When considering mechanical robustness, concepts utilising thin substrates or free standing wire grids are likely to prove difficult to manufacture with the required size. Furthermore, performance variability and device survivability is likely to be a considerable concern. The devices that fall in to this category are the air-gap mesh-HWP (M-HWP), the reflective HWP (R-HWP) and the Variable-delay Polarisation Modulator (VPM). We also mentioned that the VPM has an additional disadvantage arising from not modulating the Q and U Stokes parameters simultaneously. Note that we have not included the TPR in this list even though its present design includes a free-standing wire-grid. This is because future developments could replace the free-standing grid with a photolithographic grid that will be included in the QWP structure and thereby substantially improve the robustness of the device.

Spaceflight, and cryogenic operation, will most likely induce significant mechanical stress within the polarisation modulators, particularly during the launch phase. It is therefore reasonable to assume that options based on dielectrically embedded grids will be more robust. Consequently, at this stage, we consider as viable solutions the following (also highlighted in the table):

- Embedded mesh-HWP (EM-HWP);
- Embedded reflective-HWP (ER-HWP);
- Translational polarization rotator (TPR).

We note that the EM-HWP has immediate potential for offering a relatively lower mass as it is made essentially of plastic, whereas the other ER-HWP and TPR types require an additional metallic mirror. On the other hand, the ER-HWP (in both versions) has the advantage of possessing a reduced temperature gradient across its surface due to the contact with a metallic mirror. Common to the three pre-selected options is the additional advantage that materials required to manufacture all these devices are commercially available and the fact that they can be produced using the same photolithographic techniques. Facilities necessary for the photolithographic processes, grid production, and substrate bonding are available within the project team and are located at the Cardiff and Manchester Universities, although they allow the manufacture of embedded devices only up to ~50 cm in diameter. However, the diameter requirement for the HWP development of this work was set to 120 cm and in order to develop these large plates the collaboration is undertaking a major upgrade of their facilities. The upgrade is in progress and it will soon allow the production of 120 cm diameter mesh-embedded devices.

## 3.2 HWP RF performance

As discussed in the polarisation modulators review section, there are a considerable number of key RF performance parameters that affect and define the suitability of the polarisation modulation options. These include, for instance, signal transmission and reflection, differential transmission and reflection, cross-polarisation, bandwidth, modulation efficiency, co-polar beam impact, cross-polar beams, and homogeneity performance across the plate. In Table 2 we have summarised our knowledge of these parameters for the different design concepts, once again using the same meaning for the colours. We included all of the polarisation modulation options in order to present an overall overview of present-day status.

One of the most important parameters to study is the RF bandwidth, i.e. the frequency range within which the RF performance of the device meets operational requirements. In the COrE project proposal, to which this study is addressed, there are 15 frequency bands spanning from ~45GHz to ~800GHz. This is achieved using a free-standing reflective HWP that, for the manufacturing issues discussed above, is extremely challenging to realise. The fractional bandwidth is relatively large and multi-octave, of the order of ~180%. In order to tackle this very demanding requirement we can either try to design a device with an extremely large bandwidth or another one with multiple bands coincident with the ones of the instrument.

In the first case, very broadband performance could be in principle achieved using the Pancharatnam approach. Normally used with birefringent materials, this technique can also be applied to mesh-based devices. Both the EM-HWP and the TPR are based on mesh grids and Pancharatnam type recipes are available both for HWPs and QWPs, thereby allowing us to apply this principle to each. Antireflection coatings are required as well to optimise the broad overall transmissions. The results of our preliminary simulations and optimisations showed that bandwidths of ~100% can be achieved. Even so, although further design optimisations might lead to broader bandwidths, at this stage these options cannot be the considered as baseline option.

The second approach consists in developing dielectrically embedded versions of the free-standing reflective HWP. One solution is the ER-HWP-A option discussed in Section 2.6 (Fig.11a). This device behaves almost like a free-standing R-HWP showing the required periodic frequency bands and even in presence of dielectric substrates the performance can be kept at the same level or above that of an air-gap R-HWP. The other solution, the ER-HWP-B option (Fig.11b), based on embedded mesh artificial dielectrics, can achieve extreme large bandwidths of ~150%. Further developments might lead to meet the whole COrE requirements of an 180% wide bandwidth.

The bandwidth requirement is so defining a prioritisation between the pre-selected options that are now in the following order:

1) Embedded reflective-HWP (ER-HWP-A or -B)   → Multiple narrow periodic bands or large bandwidth
2) Embedded mesh-HWP (EM-HWP)                 → Single large bandwidth
3) Translational polarization rotator (TPR)   → Single large bandwidth

Each of the three options satisfies the requirement of modulating both the Q & U Stokes parameters and to have very high polarisation modulation efficiency. The transmission, reflection and absorption coefficients/differential-coefficients will be evaluated and constrained during the detailed design in order to meet the requirements. Concerning the ellipticity and cross-polarisation requirements, preliminary studies of EM-HWP and ER-HWP-A designs show that these parameters can be kept under control and within requirements. The flatness requirement, preliminarily investigated on small diameter devices, resulted to not be a serious concern in the case of embedded mesh devices.

Table 1. Polarisation modulators trade-off table from the manufacture point of view.

| Manufacture | Transmission modulators | | | | Reflection modulators | | | |
|---|---|---|---|---|---|---|---|---|
| Mechanical modulation | Rotation | | | | Rotation | | | Translation |
| Material | Birefringent crystal | | Metal grids | | Metal grids + Flat mirror | | | |
| Modulator type | Single Plate | Multi-Plate Pancharatnam | Air-gap Mesh-HWP | Embedded Mesh HWP | Air-gap Reflective HWP | Embedded Reflective HWP | Variab. delay Polarization Modulator (VPM) | Translational Polarization Rotator (TPR) |
| Modulator sketch | B-HWP → | BP-HWP → | M-HWP → | EM-HWP → | R-HWP → | ER-HWP → | VPM ↕ | TPR ↔ |
| Material availability | Limited to ~30 cm ⌀ (Sapphire) | Limited to ~30 cm ⌀ (Sapphire) | > 120 cm ⌀ (Cu on PP) | > 120 cm ⌀ (Cu on PP) | > 120 cm ⌀ (Cu on PP) | > 120 cm ⌀ (Cu on PP) | > 120 cm ⌀ (Cu on PP) | > 120 cm ⌀ (Cu on PP) |
| Mechanical robustness | High | High | Very Low | Medium | Very Low | High | Very Low | Medium |
| Relative Mass (without mechanism) | High (crystal) | High (crystal) | Very low (thin plastic) | Low (bulk plastic) | High (mirror) | High (mirror) | High (mirror) | High (mirror) |
| Maximum achievable diameter | ~30 cm | ~30 cm | 52 cm | 33 (52 cm) / 120 cm | 52 cm | 33 (52 cm) / 120 cm | 52 cm | 33 (52 cm) / 120 cm |

Legend: Advantage (green) | Disadvantage (grey) | Not usable (red) | Goal (yellow)

Table 2. Polarisation modulators trade-off table in terms of RF performance.

| RF performance | Transmission modulators | | | | Reflection modulators | | | |
|---|---|---|---|---|---|---|---|---|
| Mechanical modulation | Rotation | | | | Rotation | | | Translation |
| Material | Birefringent crystal | | Metal grids | | Metal grids + Flat mirror | | | |
| Modulator type | Single Plate | Multi-Plate Pancharatnam | Air-gap Mesh-HWP | **Embedded Mesh-HWP** | Air-gap Reflective HWP | **Embedded Reflective HWP** | Variab. delay Polarization Modulator (VPM) | **Translational Polarization Rotator (TPR)** |
| Modulator sketch | B-HWP → | BP-HWP → | M-HWP → | EM-HWP → | R-HWP → | ER-HWP → | VPM ↕ | TPR ↕ |
| Modulated Stokes parameters | Q & U | Q & U | Q & U | Q & U | Q & U | Q & U | Q or U | Q & U |
| Maximum bandwidth | Narrow | Very Broad ~110% | Broad ~80% | Very Broad ~100% | Narrow | Narrow (ER-A) / ~150% (ER-B) | Narrow | Narrow |
| Multiple sub-bands | Need periodic ARC | Limited to max BW | Limited to max BW | Limited to max BW | Periodic | Periodic | Variable | Variable Limited by QWP BW |
| Modulation efficiency | > 99% | > 99% | > 99% | > 99% | > 90% | > 90% | > 90% | > 90% |
| Transmission / Differential transmission | > 90 % / < 1 % | > 90 % / < 1 % | > 90 % / < 1 % | > 90 % / ~ 1-2 % | Not applicable | Not applicable | Not applicable | Not applicable |
| Reflection / Differential Reflection | 3% / 0.1% | 3% / 0.1% | < 4% / < 2% | < 4% / < 2% | > 98 % / < 1 % | > 98 % / < 1 % | > 98 % / < 1 % | > 98% / < 2% |
| On-axis Average Cross-Polarisation | <-20 dB (6% BW) | <-20 dB (110% BW) | <-20dB (80%BW) | -20dB (90% BW) / -30dB (30% BW) | -30dB (within bands) | -20dB (150% BW) | Not Available | Not available |
| Co-Polar beam impact Ellipticity | Not available | Not available | Not available | ~1% (25% BW) | < 1% | ~ 1-2% | Not available | Not available |
| Cross-Polar beams | Not available | ~ -30dB | Not available | <-35dB (25% BW) | < -30dB | <~-25dB (ER-A) / Not Av. (ER-B) | Not available | Not available |
| Flatness / Homogeneity | Very high | High | High | High (TBC) | High | High (TBC) | High | High (TBC) |

Advantage: green | Disadvantage: grey | Not usable: red

Table 3. Polarisation modulators trade-off table from the cryogenic and mechanical operation point of view.

| Cryo-mechanical | Transmission modulators | | | | Reflection modulators | | | |
|---|---|---|---|---|---|---|---|---|
| Mechanical modulation | Rotation | | | | Rotation | | | Translation |
| Material | Birefringent crystal | | Metal grids | | Metal grids + Flat mirror | | | |
| Modulator type | Single Plate | Multi-Plate Pancharatnam | Air-gap Mesh-HWP | **Embedded Mesh HWP** | Air-gap Reflective HWP | **Embedded Reflective HWP** | Variab. delay Polarization Modulator (VPM) | **Translational Polarization Rotator (TPR)** |
| Modulator sketch | B-HWP → | BP-HWP → | M-HWP → | EM-HWP → | R-HWP → | ER-HWP → | VPM | TPR ↕ |
| Modulation mechanism | Cryogenic mechanical rotator | | | | | | Cryogenic mechanical translator | |
| Mechanism cryogenic heat dissipation | Low | | | | | | Not Available | |
| Waveplate robustness to cooling cycles | Low (ARC problems) | Low (ARC problems) | High | High | High | Not Available | High | High |
| Waveplate gradient Temperature | Low | Low | Not Available | Not available Under study | Not Available | Not available Under study | Not Available | Not Available |
| Waveplate low temperature emissivity | Sapphire < 1% | Sapphire < 1% | Low | Low | Low | Very Low | Low | Low |
| Waveplate low T differential emissivity | Low | Low | Low | Low Under study | Low | Low Under study | Low | Low |

Advantage: green | Disadvantage: grey | Not usable: red

## 3.3 HWP suitability for cryogenic / mechanical operation in space

In Table 3 we have summarised the key aspects of all the HWP options from the cryo-mechanical point of view. Our discussion will continue considering just the three pre-selected options. The cryogenic modulation mechanism will be a rotator, in the case of the EM-HWP and ER-HWP, or a translator, in the case of the TPR. The key requirements of this mechanism are to have very low thermal dissipation at cryogenic temperatures (goal of 10mW at 10K) and reliable and repeatable performance.

Different cryogenic mechanisms for polarization modulators are described in the literature (see [16, 17, 18, 19]). All have advantages and disadvantages, in terms of accuracy, reliability, potential systematic effects. In particular:

- The continuous rotation WP based on magnetic bearings is optimal in terms of residual friction and modulation speed, but problematic in terms of thermal control of the WP and complex in fabrication, and requires anyway a motion transmission; moreover Eddy currents are inevitably produced in the structure, producing power dissipation.

- The stepped-rotation WP is simpler to build, allows for thermalization links, and does not require special components. It can be problematic in terms of residual friction and wear of the bearings in long-life applications like ours.

- The linear motion required for the TPR allows for thermalization links, but can be problematic in terms of residual friction, wear and parallelism / planarity of the optical components.

Given the large diameter of the HWP and TPR, a rotator is much more reliable than a translator: while precision in the motion of the former is provided by simple rotary bearings, for the latter it requires several (given the size) synchronized linear actuators. From the cryo-mechanical point of view, the ER-HWP is compatible with simpler and more reliable rotator mechanisms. The ER-HWP requires a reflective flat substrate, i.e. a metal plate. This solves very efficiently the thermalization problem affecting the central part of the EM-HWP (and TPR). Moreover, a solid metallic disk can be rotated by means of a central shaft, connecting the cold disk to a warm rotator, and acting as a central support system. In the case of the EM-HWP, instead, the center region has to be free from mechanical parts, which means that the support and rotation mechanisms must have a thoroidal shape.

Prototyping a cryogenic rotator is considerably simpler in the case of the ER-HWP. Parts of the rotator are at room-temperature and a long insulating shaft, supported by bearings at both warm and cold ends. A suitable mock-up of the ER-HWP can be built so that its diameter is smaller than that of the real HWP, while the moment of inertia and mass are the same. In this way a small cryogenic facility can accommodate all the real rotator parts and the mock-up ER-HWP. The thoroidal rotator for an EM-HWP, instead, requires a rotator diameter larger than the one of the HWP itself, i.e. of the order of 1.5m. As a consequence the cold volume of the cryostat should have an even larger diameter and the time required to complete the thermal cycles would be much longer. The control of radiation loads is also much more critical.

## 3.4 HWP baseline selection

Combining together the previous discussions about manufacturability, RF performance and cryo-mechanical operations, we can finally select our main HWP technology options:

- <u>Baseline option</u>:   Embedded Reflective Half Wave Plate (ER-HWP)

- <u>Back-up solution</u>:   Embedded Refractive Half-Wave Plate (EM-HWP)

The main reasons that drove us to these conclusions can be summarised as follows:

1. Only three out of the eight options can be manufactured: EM-HWP, ER-HWP and TPR.

2. On the RF point of view the ER-HWP can be either extremely broadband (B) or have periodic sub-bands (A) similar to the air-gap R-HWP version. The option (A) allows covering all the COrE frequency bands. The refractive EM-HWP and TPR options will require further studies to increase the bandwidth from the present ~100% to the required 180%. Although the ER-HWP works in off-axis conditions (~45° incidence angle) the induced ellipticity and cross-polarisation of the beams in our preliminary simulations resulted to be within the requirements.

3. The ER-HWP will be based on a dielectrically embedded mesh structure in contact with a thick metallic mirror. The mirror will function also as a thermal short allowing the best possible thermalisation for a waveplate. The EM-HWP does not have this advantage.

4. The ER-HWP mechanism can be much simpler than the one required for a refractive EM-HWP or a TPR and is obviously compatible with a room temperature motor and a simple shaft for motion transmission. In this case the cryogenic bearings do not need to be bigger than the waveplate, in order to include it. The mechanism can be much smaller and then interfaced to the center of the waveplate mirror structure. In addition, such mechanism will not require a large cryostat to be tested.

## 4. CONCLUSIONS

We have discussed eight different quasi-optical polarization modulators ranging from birefringent materials to metal mesh based devices and within the refractive and reflective modulator families. Working principles, advantages and disadvantages of each device were presented. This study was carried out in order to assess which modulator could be manufactured with large diameters (>1m) and be considered as the polarization modulator for a future CMB B-mode satellite mission. Keeping in mind this type of target, detailed trade-off tables have been presented in terms of manufacturability, RF performance and cryo-mechanical operation. A first selection due to the availability of materials and robustness of the devices restricted the number of options down to three. RF performance, specifically the bandwidth, and cryo-mechanical considerations led to the choice of the baseline option and back-up solution being respectively an Embedded Reflective HWP (ER-HWP) and an Embedded Mesh-HWP (EM-HWP). An ongoing facility upgrade within the team institutions will allow soon the realization of 120 cm diameter mesh based devices.

## 5. AKNOWLEDGMENTS

This work was supported by the European Space Agency under Technology Research Program contract # 4000107865/13/ML/MH. We would like to thank Dr David Chuss and Dr Giorgio Savini for the useful discussions about polarization modulation techniques.